\documentclass[twocolumn,showpacs,amsmath,amssymb,nofootinbib]{revtex4-1}
\usepackage{dcolumn}
\usepackage{bm}

\newcommand{\be}{\begin{equation}}
\newcommand{\ee}{\end{equation}}
\newcommand{\ba}{\begin{eqnarray}}
\newcommand{\ea}{\end{eqnarray}}
\newcommand{\nn}{\nonumber\\}
\newcommand{\n}[1]{\label{#1}}

\newcommand{\hh}{\, ,\hspace{0.5cm}}
\newcommand{\hhh}{\, ,\hspace{0.2cm}}
\newcommand{\bn}{{\mbox{\boldmath $\nabla $}}}
\newcommand{\BM}[1]{{\mbox{\boldmath $#1$}}}

\newcommand{\ve}{\varepsilon}

\newcommand{\bi}[1]{\bibitem{#1}}

\newcommand{\JMP}[4]{{#1}{J. Math. Phys.\ }{\bf #2},\  #3 (#4).}
\newcommand{\PRD}[4]{{#1}{Phys. Rev.\  D\ }{\bf #2},\  #3 (#4).}
\newcommand{\PRA}[4]{{#1}{Phys. Rev.\  A\ }{\bf #2},\  #3 (#4).}

\newcommand{\PRE}[4]{{#1}{Phys. Rev.\  E\ }{\bf #2},\  #3 (#4).}
\newcommand{\PR}[4]{{#1}{Phys. Rev.\ }{\bf #2},\  #3 (#4).}
\newcommand{\CMP}[4]{{#1}{Comm. Math. Phys.\ }{\bf #2},\  #3 (#4).}

\newcommand{\PRL}[4]{{#1}{Phys. Rev. Lett.\ }{\bf #2},\  #3 (#4).}
\newcommand{\GRG}[4]{{#1}{Gen. Rel. Grav.\ }{\bf #2},\  #3 (#4).}

\newcommand{\NPB}[4]{{#1}{{Nucl. Phys.}\ B\ }{\bf #2},\  #3 (#4).}
\newcommand{\AP}[4]{{#1}{Ann. Phys., N. Y.\ }{\bf #2},\ #3 (#4).}

\newcommand{\SPU}[4]{{#1}{Soviet Phys. Uspekhi\ }{\bf #2},\  #3 (#4).}

\newcommand{\HPA}[4]{{#1}{Helv. Phys. Acta\ }{\bf #2},\ #3 (#4).}
\newcommand{\OA}[4]{{#1}{Optica Applicata\ }{\bf #2},\  #3 (#4).}
\newcommand{\JPU}[4]{{#1}{J. Phys. of USSR\ }{\bf #2},\  #3 (#4).}
\newcommand{\NA}[4]{{#1}{Nature\ }{\bf #2},\  #3 (#4).}
\newcommand{\BOOKIN}[5]{{#1,\ }{in {\it #2}}\ {edited by #3\ }{(#4)\ }{p. #5.}}
\newcommand{\BOOK}[4]{{#1}{ {\it #2\ }}{(#3)\ }}

\begin{document}

\title{Spinoptics in a stationary spacetime}
\author{Valeri P. Frolov}
\email{vfrolov@ualberta.ca}
\author{Andrey A. Shoom}
\email{ashoom@ualberta.ca}
\affiliation{Theoretical Physics Institute, University of Alberta,
Edmonton, AB, Canada,  T6G 2G7}

\begin{abstract}
The main goal of the present paper is to study how polarization of photons affects their motion in a gravitational field created by a rotating massive compact object. We study propagation of the circularly polarized beams of light in a stationary gravitational field. We use (3+1)-form of the Maxwell equations to derive a master equation for the propagation of monochromatic electromagnetic waves of the frequency $\omega$ with a given helicity. We first analize its solutions in the high frequency approximation using the `standard' geometrical optics approach. After that we demonstrate how this `standard' approach can be modified in order to include the effect of the helicity of photons on their motion. Such an improved method reproduces the standard results of the geometrical optics at short distances. However, it modifies the asymptotic behavior of the circularly polarized beams in the late-time regime. We demonstrate that the corresponding equations for the circularly polarized beam can be effectively
obtained by modification of the background geometry by including a small factor proportional to $\omega^{-1}$
whose sign corresponds to photon helicity. We obtain the modified equations for circularly polarized rays by using such a `renormalization' procedure, and calculate the corresponding renormalization term for the Kerr geometry.
\end{abstract}

\pacs{41.20.Jb, 42.15.-i, 42.81.Gs, 04.20.Cv, 04.70.Bw \hfill
Alberta-Thy-10-11}

\maketitle

\section{Introduction}

The main goal of the present paper is to study how polarization of photons affects their motion in a gravitational field created by a rotating massive compact object.

It is well known  (see e.g. \cite{Pl:60,VoIzSk:70,Ma:73}) that
the electromagnetic field equations in an external gravitational
field  are formally equivalent to the Maxwell
equations in a flat spacetime written in Cartesian coordinates in the presence of  a ``medium'' with the dielectric permittivity and magnetic permeability tensors related to the spacetime metric\footnote{It is interesting to notice that the Maxwell equations in a curved spacetime filled with a moving fluid
with non-trivial dielectric permittivity and magnetic permeability can be identically rewritten as the vacuum
Maxwell equations in a specially modified so-called {\em Gordon's optical metric}. This result was obtained by Gordon \cite{Gordon}. For recent discussion and applications see, e.g., \cite{ChenKant1} and \cite{ChenKant2}.}.
This approach was used to study different aspects of propagation of
electromagnetic waves in a gravitational background
\cite{Ma:73,Bal,Ma:74,Ma:75}.  In particular, using this approach it was shown that
for the motion of a photon in the external stationary gravitational field its helicity is conserved \cite{Ma:73,Ma:75}.
This implies that in a stationary external gravitational field, in
the absence of photon creation, the state of the circular polarization remains the same.  In particular, scattering of the incoming right (left) circularly polarized  electromagnetic radiation in an asymptotically flat spacetime
results in the outgoing right (left) circularly polarized  radiation.
For such problems one can decompose an electromagnetic wave into two independent non-interacting
components of the left and right circular polarization\footnote{If a gravitational field creates photons, an initially
circularly polarized beam of photons may have in the final state created photons of different polarization. This can happen if the gravitational field is time dependent, or stationary, for example in the case when the beam is propagating through the ergosphere of a Kerr black hole, where the superradiant paticle creation is possible.}.

If the wavelength of electromagnetic waves is much smaller than
the characteristic scale of the gravitational field,  one
can use the geometrical optics approximation for their description.  In
this approximation the wave equations are reduced to the
Hamilton-Jacobi equation for the eikonal and transport equations for the amplitude of the wave. The former is equivalent to
geodesic equations for null rays. If an electromagnetic wave is
linearly polarized, then in the geometrical optics approximation the
vector of polarization is parallelly propagated along the ray (see, e.g., \cite{Ehl} and 
\cite{MTW}, p. 570). First order corrections to the geometrical optics approximation where discussed by Ehlers \cite{Ehl} and polarization effects in the Schwarzschild metric were studied in \cite{DwiKan} and \cite{Kan2}.

In a general case,  as a result of the bending of a light ray,
the polarization vector changes its direction. Besides this, if the source
of the gravitational field is rotating, the polarization vector can
rotate around the propagation vector. This effect is a gravitational
analogue of the well-known electromagnetic Faraday effect, which
appears in magneto-active media
\cite{Zo:99,Se:04}. This effect was studied in the
Kerr spacetime, where the equations of the parallel transport along null
geodesics can be solved exactly \cite{IsTaTo:88,StCo:77,CoSt:77,CoPiSt:80}. The relativistic
effect of Faraday rotation may be important for the explanation of
the properties of radiation from accreation disk of a rotating black
hole. In particular, Laor, Netzer, and Piran
demonstrated that  this effect can suppress polarization of the
optical radiation from the active galactic nuclei  \cite{LaNePi:90}.

The effect of the gravitational Faraday rotation of the spin of a photon
is the result of the action of the photon's trajectory in a given gravitomagnetic field on the spin orientation.
One can expect a dual effect, namely an action of the helicity of the photon on its trajectory.
Such an effect is analogous to the optical Magnus effect which was studied for the propagation of light in an optically inhomogeneous medium \cite{LibZel}.
In other words, when a polarized radiation propagates near a {\em rotating}
gravitating body photons of the
opposite initial helicities are scattered differently
\cite{Ma:74,Ma:74a}. It was also shown that the scattering amplitude of the left and right circularly
polarized waves are different \cite{Ma:73,Ma:75}.
 Because of these effects, one can expect a partial
polarization of the scattered by the Kerr black hole initially unpolarized
radiation \cite{Ma:73}. The helicity asymmetry due to the gravitational deflection of unpolarized light by a rotating body was calculated in \cite{Gu,BaGu}.

The dependence of photons scattering by a rotating black hole on
their helicity is an effect beyond the lowest order of the geometrical
optics approximation.
Suppose that two beams of light  of the same wavelength $\lambda$ but opposite helicity
pass near a rotating object of the mass $M$ and the angular momentum $J$ at the distance $L\gg GM/c^2$.
Then the expected  angular difference  of their asymptotic trajectories is of the
order of $GJ\lambda/(c^3 L^3)$. This estimation was obtained by Mashhoon
\cite{Ma:75,Ma:74a} in the weak field approximation for
the gravitational field of a rotating body.

The aim of this paper is to study the propagation of
circularly polarized beams of light in a stationary spacetime assuming that the gravitational field is not weak. This paper is organized as follows. In Sec. II we discuss
general aspects of the geometrical optics approximation for the polarized light propagation
in an external stationary gravitational field. This section contains references to the numerious results related to our
problem and its analogues for the spin particle motion in a magnetic field. In Sec. III
we use (3+1)-decomposition of a stationary gravitational field and derive 3-dimensional master equation for the propagation of monochromatic elecromagnetic waves with a given state of
helicity. The `standard' scheme of the geometrical optics approximation applied to the master equation is presented in Sec. IV. Section V describes a modified approximation which takes into account the back reaction of the photon polarization state on its trajectory. We also demonstrate that the corresponding modification of the photon equation of motion can be effectively obtained by simple `renormalization' of the background gravitational field, depending on the polarization of the photon and its frequency. In order to illustrate
this procedure, we calculate the corresponding `renormalization' term for the Kerr geometry. General remarks and
possible applications of the developed formalism are discussed in Sec. VI.  The Appendix contains additional technical results.

In this paper we use the system of units where $G=c=1$ and the sign conventions for the metric
and other geometrical quantities adopted in the book by Misner, Thorne,
and Wheeler \cite{MTW}.

\section{Application of the geometrical optics to the polarized light beam propagation}

Description of the light propagation by means of the geometrical optics is an approximation. This approximation is valid when the wavelength of light $\lambda$ is much smaller that other characteristic scales of the problem, such as (i) the length $l_m$ at which the parameters of the medium change significantly; (ii) the radius $l_R$ of the spacetime curvature, and (iii) the radius $l_w$ of the curvature of the wave front surface. Let $l$ be the smallest of these scales, then the condition of validity of the geometrical optics approximations is $\ve\equiv\lambda/(2\pi l) \ll 1$. Under this condition one expects that the field is fast oscillating while its amplitude changes slowly.
The mathematical version of these physical ideas are asymptotic formulas for solutions of the corresponding field equations. The method of obtaining these asymptotics is well known (under different names)  and widely used both in physics and mathematics. For example, in quantum mechanics it is called the WKB method, and the smallness parameter $\ve$ is proportional to the Planck constant $\hbar$.

The general scheme of the construction of the asymptotic stationary solutions in the geometrical optics approximations is the following (see, e.g., \cite{Arnold}):
\begin{itemize}
\item Let $F(x)$ be a field defined on an $n$-dimensional manifold with coordinates $x$. Write it in the form
$F(x)=f(x)\exp(i{\cal S}(x)/\ve)$, and substitute this ansatz into the field equation(s);
\item In the obtained relation select the leading in $\ve$ terms. A condition that the corresponding expression vanishes gives the eikonal equation, which is usually of the form
$H(\nabla {\cal S},x)=0$;
\item In order to solve this first order partial differential equation, one considers it as the Hamilton-Jacobi equation for a particle with the Hamiltonian
$H(p,x)$, where $p\equiv\nabla {\cal S}$ is the particle momentum, and the initial conditions ${\cal S}={\cal S}_{0}(x)=const.$, such that $p_0(x)=\nabla {\cal S}_0(x)$.
\item Let $\ell$ be a parameter along the particle trajectory, and let $x(\ell)|_{\ell=0}=x_{0}$, $p(\ell)|_{\ell=0}=p_{0}$ be the initial conditions.
\item Solve the Hamilton equations of motion for the initial conditions. The set of the solutions $\{x(\ell), p(\ell)\}$  determines an $n$-dimensional submanifold of the phase space, which is known as a Lagrange manifold. Its characteristic property is that the symplectic form  vanishes on it. The function ${\cal S}(x)$ is the action defined on the particle trajectories.
\item Next to the leading order in $\ve$ equations give transport equations
which allow one to find the amplitude $F$ along the particle trajectories, and hence, determine it on the Lagrange manifold. At the points where the projection of the latter on the configuration space is regular, this projection defines the required asymptotic expansion of the amplitude
$f(x)\sim f_0(x)+\ve f_1(x)+\ldots \, $.
\item Points of the Lagrange manifold where its projection on the configuration space is singular form caustics, which require special consideration (for details see, e.g., \cite{KrOr_1,KrOr_2} and the references therein).
\end{itemize}

In the application of the geometrical optics to the problem of the polarized light beam propagation several new features become important. First of all, the light amplitude has not one but several components. As a result, the transport equation for the amplitude propagation along the particle trajectories has a matrix form.  One of the methods used in such case is based on the diagonalization of the corresponding equations (see, e.g., \cite{BjoOrb,WeiLit}). A similar approach is used for the construction of the geometrical optics asymptotic solutions of the Maxwell equations in magneto-active media. In this case, after the diagonalization one has independent modes. Moreover, the eikonal equations for different modes are generally different (see, e.g., \cite{KraNaiFuj,KraBie} and the references therein). A remarkable desription of the general eikonal method for systems with several components is given in \cite{weinberg}.

Our main subject, spinoptics in a stationary gravitational field, in many aspects is similar to  more  analyzed and better understood problem of a charged spin particle motion in an external electromagnetic field.
A  well-known example of such a problem is the Stern-Gerlach experiment. Observations show that when a quantum charged particle with spin (electron)
 propagates through a region of inhomogeneous magnetic field it gets deflected differently for different spin orientations.
In order to describe this effect, one can use the WKB approximation to derive quasi-clasical equations of the electron motion. The problem is that the electron magnetic moment interacting with the gradient of the magnetic field is proportional to $\hbar$, and hence, it vanishes in the limit $\hbar\to 0$. As a result, a simple application of the WKB does not allow to obtain the required result for the spin-dependent deflection of the electron in the Stern-Gerlach experiment. This conclusion made by Pauli \cite{Pauli} and supported by more accurate calculations \cite{Galanin,RubKel,RafSch} stimulated many interesting discussions (see, e.g., the review \cite{KraNaiFuj} and the references therein).

A general resolution of this puzzle is the following. In the Stern-Gerlach experiment one deals with a scattering problem.
That is why in the quasi-classical description one needs to know the WKB asymptotic solution not only locally, but globally as well.
The spin-dependent terms, which are small (of the order of $\hbar$), locally give a small contribution to the electron trajectory. But as a result of the accumulation
of these spin-dependent corrections during the scattering, their effect on the electron trajectory at the later time is not small at all.
In order to be able to describe this effect and to construct the corresponding global WKB solution one needs to include the lowest order spin-dependent corrections in the eikonal equation. This results in the global modification of the corresponding Lagrange manifold. One can expect that this modification is locally small, but becomes important in the late-time regime.
This approach was originally advocated by de Broglie \cite{Bro}. It was
made more concrete ten years later by Schiller \cite{sch_1,sch_2}.

Several different approaches to the problem of motion of a spin 1/2 particle  in a magnetic field in the WKB approximation have been developed recently. The most consistent approach is to include spin degrees of freedom into the quasi-classical equations (see, e.g., \cite{BolKep,Kep,Bol}). As a result, the phase space of the particle becomes a direct product the standard $2n$-dimensional phase space and a two-dimensional sphere. The Hamiltonian defined on this enlarged phase space describes both the spin rotation induced by the electron motion in the magnetic field and the back reaction of the spin rotation on the
particle motion \cite{Bol}. In this scheme, the symplectic form is also modified and is determined on the enlarged phase space \cite{Duv}. A similar geometrical approach to  a theory of  spinoptics in refractive media can be found in \cite{DHH1} and \cite{DHH2}.

Another way to include spin-field interaction is to modify the eikonal by including into it a term dependent on the spin orientation, which is proportional to $\hbar$ \cite{BlBlSaNo}.
This modification does not change the local WKB expansion of the field and can be considered as its trivial redefinition. However, these locally small corrections of the eikonal equation result in the non-trivial change of the late-time asymptotics of the particle trajectories. A similar result was obtained in \cite{BliBli} where there was formulated a general principle which can be used for identification of the terms which must be included into the action ${\cal S}$. The authors called this method the {\em modified geometrical optics}.

An analogical effect of the photon spin on its motion is known as the spin-Hall effect \cite{BlNiKlHa,BlGoKlHa,OnMuNa_1,OnMuNa_2,SuOnMuNa}.
This intrinsic spin-Hall effect, associated with the Berry phase, was observed experimentally for the propagation of helical light beam at a grazing angle inside a glass cylinder \cite{BlNiKlHa}. The Hamiltonian and Lagrangian approaches to the problem of evolution
of spinning light in an inhomogeneous medium are summarized,
respectively, in the papers \cite{BFK2007} and \cite{B2009}.

In our approach to the spinoptics in a stationary gravitational field, we demonstrate how the next to the leading order  helicity-dependent corrections  naturally
arise in the eikonal equations and develop the corresponding improved scheme, which allows one to describe the effect of the state of circular polarization of the light beam on its motion. We use the similarity of this problem to the better-understood case of the spin particle motion in a magnetic field. In spite of this similarity, the case of light motion has  one essential difference: photons are massless while electrons have mass. In the quantum description of a photon, its state can be characterized by the photon's helicity. The spin vector of massive particles can have an arbitrary orientation with respect to their velocity vector. In the classical description of light, a state with a given helicity corresponds to either right of left circular polarization of the electromagnetic wave. The photon's helicity is a conserved quantity \cite{Ma:73,Ma:75}. If an external gravitational field does not create photons, the number of photons in a beam and their helicity remain the same in the scattering process. In other words, solutions of the source-free Maxwell equations in an external gravitational field can be separated into the sum of two solutions, one with right and another with left circular polarization. In a general case, this decomposition involves self-dual and anti-self-dual complex fields \cite{Cas3}. Thus, to study the spinoptics effects in a gravitational field, it is natural to consider electromagnetic waves with a given circular polarization. In our approach, we write such decoupled wave equations and apply the geometrical optics approximation to each of the two sectors corresponding to the right and left circularly polarized waves.

However, there is some problem remains. Namely, the Maxwell equations form a set of eight first order partial differential equations for six field variables, which are components of the electromagnetic field tensor $F_{\mu\nu}$. This redundancy is the consequence of the gauge invariance of electromagnetic field. Using (3+1)-decomposition we write the complete
set of the Maxwell equations in an equivalent form of the master equation. The master equation follows from the  set of three first order partial differential equations for 3-dimensional complex vectors of the field components with a given state of polarization. The form of the master equation is used to develop the scheme of the geometrical optics approximation. We shall also demonstrate that  the scheme is convenient and provides a natural starting point for the new improved scheme, which is the purpose of our work.

\section{Propagation of circularly polarized light in a stationary spacetime}
\label{s2}
\setcounter{equation}0

In this section we consider the propagation of a weak electromagnetic field
in a stationary gravitational field background. In order to fix our notations we collect the required equations.

\subsection{(3+1)-decomposition of a stationary metric}

A stationary spacetime possesses timelike Killing vector  $\xi^{\mu}$ generating the
time-symmetry transformation. The integral lines of this vector field
obey the equation $dx^{\mu}/dt=\xi^{\mu}$. We use the Killing time $t$
as a parameter along the integral lines of $\xi^{\mu}$ and introduce three other
coordinates $x^i$, $i=1,2,3$, which are constant along the lines.
The origin $t=0$ of the Killing time on each of the lines  is
arbitrary. Thus, there exists an arbitrariness in the choice of
$t$:
\be \n{A.56}
t\rightarrow\tilde{t}=t+q(x^i)\, ,
\ee
where $q(x^{i})$ is some differentiable function of the coordinates $x^i$.
One can choose the Killing time $t$ and the coordinates $x^{i}$
labeling the Killing trajectories as spacetime coordinates ($t\equiv x^{0},x^i$).

In these coordinates the spacetime metric can be presented in the following form:
\be\n{A.57}
ds^2=g_{\mu\nu}dx^{\mu} dx^{\nu}=-h(dt-g_i\,dx^i)^2+h\gamma_{ij}\,dx^i\,dx^j \,,
\ee
where $h=-\xi^2>0$. In what follows, all the operations involving the Latin indices $i,j, \ldots$ are
performed by using the metric $\gamma_{ij}$ and its inverse
$\gamma^{ij}$. In particular, the covariant derivative in this metric
is denoted as $\nabla_i(...)\equiv(...)_{:i}$. We preserve the notation
$(\ldots)_{;\mu}$ for the 4-dimensional covariant derivative defined with respect to the metric $g_{\mu\nu}$.

The Killing
equation $\xi_{\mu;\nu}+\xi_{\nu;\mu}=0$ implies that the metric functions in (\ref{A.57})
are time-independent. Under the coordinate transformation
(\ref{A.56}) $h$ and $\gamma_{ij}$ remain invariant, while $g_i$'s
transform as
\be \n{A.58}
g_i\rightarrow\tilde{g}_i=g_i+q_{,i}\, .
\ee
It is possible to use the gauge transformation (\ref{A.56}) to put
$\tilde{g}^i_{\ :i}=0$.
To find this transformation it is sufficient to use $q(x^{i})$, which is a
solution of the equation $q_{\ :i}^{:i}=-g^i_{\ :i}$.

The metric tensor $\gamma_{\mu\nu}$, which in a choosen coordinate system has the
components $\gamma_{\mu\nu}=\delta^i_{\mu}\delta^j_{\nu} \gamma_{ij}$,
is proportional to the projector onto a surface orthogonal to
$\xi^{\mu}=\delta^\mu_{~0}$,
\be\n{2.0}
\gamma_{\mu\nu}=h^{-1}\left(g_{\mu\nu}-{\xi_{\mu}\,\xi_{\nu}\over{\xi^2}}\right)\, .
\ee
In  a static spacetime, i.e., when $g_{i}=0$,  the metric $\gamma_{ij}$ reduces to the so-called optical metric (see, e.g., \cite{SA,Cas_1,Caso_2} and the references therein). Spatial trajectories of light rays are geodesics of the optical metric.

The metric functions in the (3+1)-decomposition of the
metric (\ref{A.57}) are related to the 4-dimensional metric
functions $g_{\mu\nu}$ as follows:
\be\n{2.1}
h=-g_{00}\, ,\hspace{0.2cm}
g_i=\frac{g_{0i}}{h}\, ,\hspace{0.2cm}-g=h^4\gamma\, ,
\ee
\be\n{2.1a}
\gamma_{ij}=-\frac{g_{ij}}{g_{00}}+{g_{0i}g_{0j}\over g_{00}^2}=\frac{g_{ij}}{h}+g_i g_j\, .
\ee
Here $g\equiv \mbox{det}(g_{\mu\nu})$ and $\gamma\equiv \mbox{det}(\gamma_{ij})$.
Let us denote $g^i\equiv\gamma^{ij}g_j$. Then using the relation
$g_{\mu\nu}g^{\nu\lambda}=\delta_{\mu}^{\lambda}$ one obtains
\be\n{2.2}
\gamma^{ij}=hg^{ij}\hh g^i=hg^{0i}\hh g^{00}=-h^{-1}(1-g_i g^i)\, .
\ee

\subsection{Maxwell equations in (3+1)-form in a stationary spacetime}

Consider a test electromagnetic field in a stationary spacetime. In
accordance to the (3+1)-decomposition of the metric (\ref{A.57}) one can also make
a similar decomposition for the components of the electromagnetic
field. Following \cite{Cas_1,Caso_2} we denote
\be
E_i\equiv F_{i0}\, ,\hspace{0.5cm}B_{ij}\equiv F_{ij}\, ,
\ee
\be
D^i\equiv h^2F^{0i}\, ,\hspace{0.5cm}H^{ij}\equiv h^2F^{ij}\, .
\ee
These objects are not independent. Using the relations
\be
F_{i0}=g_{i\mu}g_{0\nu} F^{\mu\nu}\, ,\hspace{0.5cm}
F^{ij}=g^{i\mu} g^{j\nu} F_{\mu\nu}
\ee
we derive
\be\n{D}
D_i=E_i -H_{ij}g^j \hh
B^{ij}=H^{ij}-E^i g^j+E^j g^i\, .
\ee

Let $\epsilon_{ijk}$ and $\epsilon^{ijk}$ be 3-dimensional completely antisymmetric Levi-Civita symbols
normalized by the conditions $\epsilon_{123}=1$ and $\epsilon^{123}=1$, and
\be
e_{ijk}\equiv \sqrt{\gamma} \epsilon_{ijk}\hhh
e^{ijk}\equiv{1\over \sqrt{\gamma}} \epsilon^{ijk}\hhh
\ee
be the corresponding completely antisymmetric tensors. These tensors obey the relations
\be
e_{ijk:l}=0\hh e^{ijk}_{\ \ \ :l}=0\, .
\ee
We introduce the following vectors:
\be
B^i\equiv{1\over 2}e^{ijk}B_{jk}\hh
H_i\equiv{1\over 2}e_{ijk}H^{jk}\, .
\ee
Using the relation
\be
e_{ijk} e^{lmk}=\delta^l_i\delta^m_j-\delta^m_i \delta^l_j\hhh
\ee
we obtain
\be
B_{ij}=e_{ijk} B^k\hh H^{ij}=e^{ijk}H_k\, .
\ee
To define a vector product of vectors $\BM{A}$ and $\BM{B}$ in a 3-dimensional space
we shall use the following notation:
\be
\BM{C}=[\BM{A}\times\BM{B}]\, ,
\ee
where
\be
C^{i}=e^{ijk} A_j B_k\, .
\ee
For example,
\be
[\BM{H}\times \BM{g}]^i={1\over 2}e^{ijk} e_{jlm}H^{lm} g_k=H^{ik}g_k\, .
\ee
Thus the relations (\ref{D})  take the form
\be\n{rel}
\BM{D}=\BM{E}-[\BM{g}\times \BM{H}]\hh
\BM{B}=\BM{H}+[\BM{g}\times \BM{E}]\, .
\ee
We shall introduce the following operations defined in a 3-dimensional space:
\ba
&&\mbox{div} \BM{A}\equiv (\bn, \BM{A})=A^i_{~:i}={1\over \sqrt{\gamma}}(\sqrt{\gamma} A^i)_{,i}\hhh\\
&&(\mbox{curl} \BM{A})^i\equiv [\bn \times A]^i=e^{ijk}A_{k:j}\,.
\ea
It is easy to check that the following relations are valid:
\ba
\mbox{div}(f\BM{A})&=&(\bn f,\BM{A})+f\mbox{div} \BM{A}\,,\n{r1}\\
\mbox{curl}(f\BM{A})&=&[\bn f\times\BM{A}]+f\mbox{curl}\BM{A}\,,\n{r1.2}\\
\mbox{curl}[\BM{A}\times \BM{B}]&=&\BM{A}\mbox{div}\BM{B}-\BM{B}\mbox{div}\BM{A}\nn
&+&(\BM{B},\bn)\BM{A}-(\BM{A},\bn)\BM{B}\,,\n{r1.3}\\
\bn(\BM{A},\BM{B})&=&[\BM{A}\times\mbox{curl}\BM{B}]+[\BM{B}\times\mbox{curl}\BM{A}]\nn
&+&(\BM{A},\bn)\BM{B}+(\BM{B},\bn)\BM{A}\,,\n{r1.4}\\
\mbox{div}[\BM{A}\times\BM{B}]&=&(\BM{B},\mbox{curl}\BM{A})-(\BM{A},\mbox{curl}\BM{B})\,,\n{r1.5}\\
\mbox{div}\,\mbox{curl}\BM{A}&=&0\,.\n{r1.7}
\ea
To prove the last relation we notice that its left hand side can be
written as
\be
e^{ijk}\nabla_i\nabla_j A_k=e^{ijk} R_{ijkl} A^l\, .
\ee
Since the Riemann curvature tensor obeys the
identity $R_{[ijk]l}=0$, this expression vanishes. The relations (\ref{r1})-(\ref{r1.7}) in a
curved 3-dimensional space with the metric $\gamma_{ij}$ have the same
form as the corresponding relations in a flat 3-dimensional space.

Let us now present the source-free Maxwell equations in a $(3+1)$-form.
The first set of the Maxwell equations
\be
F_{[\mu\nu,\lambda]}=0
\ee
in a stationary spacetime gives
\be\n{dB}
\mbox{div} \BM{B}=0\hh
\mbox{curl} \BM{E}=-\dot{\BM{B}}\, .
\ee
Here we denote $\dot{A}\equiv A_{,t}$.

The second set of the Maxwell
equations
\be
F^{\mu\nu}_{\ \ ;\nu}={1\over \sqrt{-g}}(\sqrt{-g} F^{\mu\nu})_{,\nu}=0\,,
\ee
in a stationany spacetime gives
\be\n{dD}
\mbox{div} \BM{D}=0\hh
\mbox{curl}\BM{H}=\dot{\BM{D}}\, .
\ee
The energy-momentum tensor of an electromagnetic field
\be\n{Te1}
T_{\mu\nu}={1\over {4\pi}}\left(F_{\mu\nu}F_{\nu}^{~\lambda}-\frac{1}{4}g_{\mu\nu}F_{\lambda\sigma}F^{\lambda\sigma}\right)\,,
\ee
satisfies the relation
\be\n{Te2}
T^{\mu\nu}_{\,\,\,\,\,\,\,;\nu}=0\,.
\ee
Thus, for  the Killing vector field $\xi^{\mu}=\delta^{\mu}_{\,\,\,0}$ we have
\be\n{Te3}
(\xi_\mu T^{\mu\nu})_{;\nu}={1\over \sqrt{-g}}(\sqrt{-g}\xi_{\mu}T^{\mu\nu})_{,\nu}=0\,.
\ee
This relation implies
\be\n{Te4}
\dot{{\cal E}}+\mbox{div}\BM{\cal V}=0\,,
\ee
where
\be
{\cal E}\equiv \frac{1}{8\pi}[(\BM{E},\BM{D})+(\BM{B},\BM{H})]\,,
\ee
is analogical to the electromagnetic energy density, and
\be\n{Poy}
\BM{\cal V}\equiv \frac{1}{4\pi}[\BM{E}\times \BM{H}]\,,
\ee
is analogical to the Poynting vector.

\subsection{Master equations for circularly polirized light}

The system of equations (\ref{dB}) and (\ref{dD}) can be rewritten in a
complex form. For this purpose we introduce the following complex quantities:
\be\n{26}
\BM{F}^{\pm}\equiv\BM{E}\pm i\BM{H}\hh
\BM{G}^{\pm}\equiv\BM{D}\pm i\BM{B}\,.
\ee
In these notations the Maxwell equations (\ref{dB}) and (\ref{dD}), and
the relations (\ref{rel}) take the form
\be\n{27}
\mbox{div} \BM{G}^{\pm}=0\hh
\mbox{curl} \BM{F}^{\pm}=\pm i \dot{\BM{G}}^{\pm}\, ,
\ee
\be\n{29}
\BM{G}^{\pm}={ \BM{F}^{\pm}}\pm i[\BM{g}\times {\bf F}^{\pm}]\, .
\ee

In a stationary space a monochromatic wave can be written as
\be
\BM{E}=e^{-i\omega t} \BM{\cal E}+ e^{i\omega t} \bar{\BM{\cal E}}
\hh
\BM{H}=e^{-i\omega t} \BM{\cal H}+ e^{i\omega t} \bar{\BM{\cal H}}\,,
\ee
\be
\BM{D}=e^{-i\omega t} \BM{\cal D}+ e^{i\omega t} \bar{\BM{\cal D}}
\hh
\BM{B}=e^{-i\omega t} \BM{\cal B}+ e^{i\omega t} \bar{\BM{\cal B}}\,,
\ee
where the bars over the vectors denote complex conjugate.
Substituting these relations into Eq. (\ref{26}) we derive
\be
\BM{F}^{\pm}=e^{-i\omega t} \BM{\cal F}^{\pm}+ e^{i\omega t}
\bar{\BM{\cal F}}^{\mp}\, ,
\ee
\be
\BM{G}^{\pm}=e^{-i\omega t} \BM{\cal G}^{\pm}+ e^{i\omega t}
\bar{\BM{\cal G}}^{\mp}\,,
\ee
where
\be
\BM{\cal F}^{\pm}=\BM{\cal E}\pm i\BM{\cal H}
\hh
\BM{\cal G}^{\pm}=\BM{\cal D}\pm i\BM{\cal B}\, .
\ee
The complex positive frequency amplitudes $\BM{\cal F}^{\pm}$ and
$\BM{G}^{\pm}$ describe left (for $-$) and right (for $+$)
circularly polarized waves. In other words, for the right (left) circularly
polarized monochromatic radiation the quantities with the label $-$
($+$) vanish.

For the monochromatic waves the system of Eqs.
(\ref{27})--(\ref{29}) reduces to
\be\n{27a}
\mbox{div} \BM{\cal G}^{\pm}=0\, ,
\ee
\be \n{28a}
\mbox{curl} \BM{\cal F}^{\pm}=\pm \omega \BM{\cal G}^{\pm}\, ,
\ee
\be\n{29a}
\BM{\cal G}^{\pm}=
{\BM{\cal F}^{\pm}}\pm i[\BM{g}\times {\BM{\cal F}}^{\pm}]\, .
\ee
Equation  (\ref{27a}) is not independent. It
follows from Eq.  (\ref{28a}) and the identity (\ref{r1.7}).

Equations (\ref{28a}) and (\ref{29a}) imply the ``master equation''
\be\n{rotP}
\mbox{curl}\BM{\cal F}^{\pm}=\pm {\omega} \BM{\cal F}^{\pm}
+i\omega  [\BM{g}\times {\BM{\cal F}}^{\pm}]\, .
\ee
Solving this master equation for $\BM{\cal F}^{\pm}$, we can find $\BM{\cal G}^{\pm}$ from Eq. (\ref{28a}), and Eq. (\ref{27a}) is satisfied automatically. The form of the master equation implies that the left and right circularly polarized waves are decoupled.

\section{`Standard' geometrical optics approximation}

\subsection{Eikonal ansatz}

Let us briefly describe now the `standard'geometrical optics in its 3-dimensional form is adopted to circularly  polarized light propagation.
Our main assumption is that the characteristic wavelength $\lambda=2\pi\omega^{-1}$ is much smaller than any other characteristic length scale $l$ of the problem. We shall use $\omega^{-1}$ as a small parameter. In fact, it always enters the expressions in the dimensionless form $\ve=(\omega l)^{-1}$. In what follows, we shall use $\omega^{-1}$ as a small parameter in our geometrical optics expansion.  Following to the geometrical optics construction (see, e.g., \cite{Ehl} and \cite{MTW}, p. 570), we split a monochromatic wave into a rapidly changing phase and slowly changing amplitude. Such a split corresponds to the following eikonal ansatz:
\be\n{E2}
\BM{\cal F}^{\sigma}=\BM{f}^{\sigma} e^{i \omega {\cal S}}\,.
\ee
Here $\sigma=\pm1$ defines polarization of the wave, as defined above, $\BM{f}^{\sigma}$ is the complex valued amplitude, and ${\cal S}$ is the real valued phase.
We shall always consider a monochromatic wave of a fixed polarization, either $\sigma=+1$ or $\sigma=-1$, and skip for brevity the superscript $\sigma$, which we shall restore in the final expressions.

Substituting the eikonal ansatz (\ref{E2}) into Eq. (\ref{rotP}) and using Eq. (\ref{r1.2}) we derive
\be\n{E3}
L\BM{f}=\sigma\omega^{-1} \mbox{curl}\BM{f}\, .
\ee
Here $L$ is a linear operator acting in a 3-dimensional complex linear vector space,
\be\n{E4}
L\BM{f}\equiv \BM{f}-i\sigma [\BM{n}\times \BM{f}]\,,
\ee
where $\BM{n}\equiv \BM{p}-\BM{g}$ and $\BM{p}\equiv \bn {\cal S}$. The vector $\omega \BM{p}$ has the meaning of the usual wave vector.

According to the standard prescription, the slowly changing complex amplitude $\BM{f}$ can be expanded in inverse powers of $\omega$ as follows:
\be\n{E4.1}
\BM{f}=\BM{f}_{0}+\omega^{-1}\BM{f}_{1}+\omega^{-2}\BM{f}_{2}+...\,.
\ee
Here $\BM{f}_{0}$ is the leading order term in the geometrical optics approximation, and the higher order expansion coefficients $\BM{f}_{i}$'s, $i\geq1$ define post-geometrical optics corrections. Substituting the expansion (\ref{E4.1}) into Eq. (\ref{E3}) we obtain the following equation:
\ba
L\BM{f}_{0}&+& \omega^{-1}[L\BM{f}_{1}-\sigma\mbox{curl}\BM{f}_{0}]\nn
&+&\omega^{-2}[L\BM{f}_{2}-\sigma\mbox{curl}\BM{f}_{1}]+\ldots =0\, .\n{E4.1b}
\ea

\subsection{Properties of the operator $L$}

The leading order equation in the high frequency limit is
\be\n{E4.2}
L\BM{f}_{0}=0\,.
\ee
In an explicit form this equation reads
\be
L^{i}_{\,\,j}f_0^{j}=\left[\delta^{i}_{\,\,j}-i\sigma e^{i}_{\,\,kj}n^{k}\right]f^{j}_{0}=0\, .
\ee
In order to have a non-trivial solution $\BM{f}_{0}$, determinant of $L$ must vanish. The corresponding relation gives
\be\n{E4.7}
(\BM{n},\BM{n})=1\, .
\ee
This is the {\em eikonal equation}. It is easy to check that $L\BM{n}=\BM{n}$. That is, $\BM{n}$ is an eigenvector
of $L$ with the eigenvalue 1.

To determine other eigenvalues of $L$ it is sufficient to find solutions of the characteristic equation
\be\n{E4.5}
\mbox{det}[(1-\lambda)\delta^{i}_{\,\,j}-i\sigma e^{i}_{\,\,kj}n^{k}]=0\,.
\ee
Calculating the determinant (see Appendix A) one obtains
the following equation:
\be\n{E4.6}
(1-\lambda)[(1-\lambda)^{2}-(\BM{n},\BM{n})]=0\,.
\ee
Using Eq. (\ref{E4.7}) one finds that the other two eigenvalues are equal to $0$ and $2$.

To find the eigenvectors for $\lambda=0,2$
let us consider two unit real vectors $\BM{e}_{1}$ and $\BM{e}_{2}$ such that
\ba
&&\hspace{-0.5cm}(\BM{e}_{i},\BM{e}_{i})=1\hhh (\BM{e}_{1},\BM{e}_{2})=(\BM{e}_{i},\BM{n})=0\hhh i=1,2\,,\n{E4.8a}\\
&&\hspace{-0.5cm}[\BM{e}_{1}\times\BM{e}_{2}]=\BM{n}\hhh [\BM{n}\times\BM{e}_{1}]=\BM{e}_{2}\hhh [\BM{n}\times\BM{e}_{2}]=-\BM{e}_{1}\,.\n{E4.8b}
\ea
The set of vectors $\{\BM{n}, \BM{e}_{1},\BM{e}_{2}\}$ forms a right-oriented orthonormal basis in a 3-dimensional linear vector space. Let us now define the following vectors:
\be\n{E4.9}
\BM{m}\equiv\frac{1}{\sqrt{2}}(\BM{e}_{1}+i\sigma\BM{e}_{2})\hhh
\bar{\BM{m}}\equiv\frac{1}{\sqrt{2}}(\BM{e}_{1}-i\sigma\BM{e}_{2})\,.
\ee
The vector $\bar{\BM{m}}$ is the complex conjugate of $\BM{m}$. According to Eqs. (\ref{E4.8a}) and (\ref{E4.8b}), these vectors satisfy the following relations:
\ba
&&(\BM{m},\BM{m})=(\BM{m},\BM{n})=0\hhh (\BM{m},\bar{\BM{m}})=1\,,\n{E4.10a}\\
&&[\BM{m}\times\bar{\BM{m}}]=-i\sigma\BM{n}\hhh [\BM{n}\times\BM{m}]=-i\sigma\BM{m}\,.\n{E4.10b}
\ea
The set of vectors $\{ \BM{n},\BM{m},\bar{\BM{m}}\}$ forms a basis in a 3-dimensional complex linear vector space. One can check that these vectors are the eigenvectors of $L$ which correspond to the eigenvalues $\{1,0,2\}$ respectively, and the operator $L$ in this basis has the form
\be\n{E4.11}
L=
\begin{Vmatrix}
1 & 0 & 0 \\
0 & 0 & 0 \\
0 & 0 & 2
\end{Vmatrix}\,.
\ee
As expected, the operator $L$ is degenerate and its matrix  rank is 2.

Using the expression (\ref{E4.11}) it is easy to see that the solution of Eq. (\ref{E4.2}) is
\be\n{E4.15}
\BM{f}_{0}=\alpha\BM{m}\, ,
\ee
where $\alpha$ is an arbitrary complex quantity. Thus, one has
\be\n{E4.16}
(\BM{n},\BM{f}_{0})=0\, .
\ee
The expressions (\ref{Poy}) and (\ref{E4.10b}) imply that the vector $\BM{n}$ is parallel to the Poynting vector $\BM{\cal V}$ in the geometrical optics approximation.

It is easy to check that because of the degeneracy of the operator $L$, the following relation is valid
for an arbitrary vector $\BM{a}$:
\be\n{E4.13}
L\BM{a}=\BM{n}(\BM{n},L\BM{a})-i\sigma[\BM{n}\times L\BM{a}]\,.
\ee
This relation shows that Eq. (\ref{E3}) has a solution only if its right hand side obeys the linear constraint
\be\n{E4.14}
\mbox{curl}\BM{f}=\BM{n}(\BM{n},\mbox{curl}\BM{f})-i\sigma[\BM{n}\times \mbox{curl}\BM{f}]\,.
\ee
This constraint is exact.

\subsection{Ray trajectories}

The `standard' eikonal Eq. (\ref{E4.7})
\be\n{E7}
(\bn{\cal S}-\BM{g})^2=1
\ee
does not depend on the state of polarization of the wave. In fact, one can see that it coincides
with the eikonal equation for the massless scalar field $\varphi$, which obeys the equation $\Box \varphi=0$.

Equation (\ref{E7}) has the form of the Hamilton-Jacobi equation.
To solve it one uses solutions of a dynamical system defined by the following Hamiltonian:
\be\n{h4}
{\cal H}(x^i,p_i)\equiv\frac{1}{2}(\BM{p}-\BM{g})^2= \frac{1}{2}\gamma^{ij}(p_{i}-g_{i})(p_{j}-g_{j})\, ,
\ee
and the canonical symplectic form
\be
\Omega=\sum_{i=1}^3 dp_i\wedge dx^i\, .
\ee
The corresponding Hamilton equations are
\ba
\frac{dx^{i}}{d\ell}&=&{\partial {{\cal H}} \over {\partial p_{i}}}=\gamma^{ij}(p_{j}-g_{j})\,,\n{h5.1}\\
\frac{dp_{i}}{d\ell}&=&-{\partial {{\cal H}} \over {\partial x^{i}}}=\gamma^{jl}g_{l,i}(p_{j}-g_{j})\nonumber\\
&-&{1\over 2} \gamma^{jl}_{\,\,\,,i}(p_{j}-g_{j})(p_{l}-g_{l})\,.\n{h5.2}
\ea
The value of the Hamiltonian is constant on the ray trajectories. We choose
\be
{\cal H}=\frac{1}{2}\, .
\ee
This condition determines the choice of the parameter $\ell$ along the trajectories. One has
\be\n{h6}
\left({{d\BM{x}}\over {d\ell}}\right)^2=1\,.
\ee
Hence $\ell$ is the proper distance defined along an optical ray trajectory in the metric $\gamma_{ij}$.

It is easy to show that the Hamilton equations (\ref{h5.1})--(\ref{h5.2}) are equivalent to the
following second order equation:
\be\n{h8}
{D^2\BM{x}\over d\ell^2}=\left[\frac{d\BM{x}}{d\ell}\times\mbox{curl}\,\BM{g}\right]\,,
\ee
where $D/d\ell$ is the covariant derivative defined with respect to the metric $\gamma_{ij}$ along the trajectory.

Let $\Phi(x)=0$ be an equation of a 2-dimensional surface $\Gamma$ which defines the initial position of the wave front. Denote
$p^{(0)}_i\equiv\nabla_i\Phi(x)$. For each initial point $x_0^i$ of $\Gamma$ and the corresponding value $p^{(0)}_i$
one can solve  the Hamilton equations (\ref{h5.1})--(\ref{h5.2}). The solutions define ray trajectories which form a 3-dimensional
subspace of the 6-dimensional phase space called a Lagrange manifold. Its characteristic property is that the symplectic form restricted to this subspace vanishes. The action ${\cal S}$ calculated on these trajectories is a function of the final point $\BM{x}$,
\ba\n{h3}
{\cal S}(\BM{x})&=&
\int_{\BM{x}_0}^{\BM{x}}(\BM{p},d\BM{x})\nn
&=&\int_{\BM{x}_0}^{\BM{x}}\left[1+\left(\BM{g},\frac{d\BM{x}}{d\ell}\right)\right]d\ell\, .
\ea
In the 3-dimensional configuration space such 2-dimensional family of rays covers a 3-dimensional  region. We assume that through each point $\BM{x}$ passes only one ray of this family, so that the expression
(\ref{h3}) defines the function ${\cal S}(\BM{x})$. This function is the required solution of the Hamilton-Jacobi equation (\ref{E7}) (see, e.g., \cite{Arnold,KrOr_1,KrOr_2}).

Before we discuss properties of the amplitude $\BM{f}$ let us make two useful remarks.
\begin{enumerate}
\item Equation (\ref{h8}) is analogical to the equation of motion of a nonrelativistic particle of the unit charge-to-mass ratio in a 3-dimensional space in the presence of a magnetic field defined by the vector potential $\BM{g}$.
\item This equaton is nothing but a $(3+1)$-dimensional form of the 4-dimensional geodesic equation for a null ray in a stationary spacetime
\be\n{h9}
{D^2x^\mu \over d\lambda^2}=0\hhh x^{\mu}=(t,x^{i})\,,
\ee
where $\lambda$ is an affine parameter along the null ray, and $D/d\lambda$ is the covariant derivative defined along trajectory of the null ray. Equation (\ref{h8}) can also be derived from (3+1)-decomposition of the geodesic equation (\ref{h9}) for null rays (see, e.g., \cite{Cas_1}).
\end{enumerate}

\subsection{Transport equations}

To derive the transport equations for the complex amplitude $\BM{f}$ we apply the operator $\mbox{curl}$ to Eq. (\ref{E3}). Using Eq. (\ref{E4}) we derive
\be\n{f1}
\mbox{curl}\BM{f}-i\sigma\mbox{curl}[\BM{n}\times\BM{f}]=\omega^{-1}\sigma\mbox{curl}\,\mbox{curl}\BM{f}\,.
\ee
If one substitutes the expansion (\ref{E4.1}) into this equation the leading order term gives the following relation:
\be\n{f1a}
\mbox{curl}\BM{f}_{0}=i\sigma\mbox{curl}[\BM{n}\times\BM{f}_{0}]\,.
\ee
This transport equation defines the complex amplitude $\BM{f}_{0}$. Similar equations can be written for  the higher order coefficients  $\BM{f}_{i>1}$'s.

Using the properties of the operator $L$ we constructed a local basis $\{\BM{n},\BM{m},\bar{\BM{m}}\}$. The vector $\BM{m}$ is defined up to the transformation $\BM{m}\to \exp(i\psi)\BM{m}$, where $\psi$ is a function of $\BM{x}$. Let us choose the basis on the surface of the initial position of the wave front $\Gamma$ and define it along the ray trajectories by the condition that it is  Fermi propagated along the tangent to the rays vector $\BM{n}$. Denote
$\BM{w}\equiv\nabla_{\bf n}\BM{n}$, then the Fermi derivative of a vector $\BM{a}$ along a unit vector field $\BM{n}$ is
\be
{\cal F}_{\bf n}\BM{a}\equiv\nabla_{\bf n}\BM{a}-(\BM{n},\BM{a}) \BM{w}+(\BM{w},\BM{a}) \BM{n}\, .
\ee
One has ${\cal F}_{\BM{n}}\BM{n}=0$, and
the scalar product of any two Fermi propagated vectors is constant. Since $(\BM{m},\BM{n})=0$ at the initial surface, this orthogonality condition is valid along ray trajectories. Thus, the condition that $\BM{m}$ is Fermi propagated takes the form
\be\n{gfc}
\nabla_{\bf n}\BM{m}=-(\BM{w},\BM{m}) \BM{n}\,.
\ee
This condition fixes gauge ambiguity in the choice of the complex basis $\{\BM{n},\BM{m},\bar{\BM{m}}\}$. We call such a choice of the basic vectors fields canonical.

Denote $f_{0}\equiv (\bar{\BM{f}}_{0},\BM{f}_{0})^{1/2}$, then the complex polarization vector in the canonical frame takes the form
\be\n{f7}
\BM{f}_{0}=f_{0}\BM{\pi}_{0}\hh
\BM{\pi}_{0}=\BM{m}e^{i\varphi}\,.
\ee
To find the amplitude $f_{0}$ and the phase $\varphi$ we proceed as follows.
Using Eq. (\ref{r1.3}) we present Eq. (\ref{f1a}) in the form
\be\n{f2}
\mbox{curl}\BM{f}_{0}=i\sigma\left[\BM{n}\,\mbox{div}\BM{f}_{0}-\BM{f}_{0}\mbox{div}\,\BM{n}+(\BM{f}_{0},\bn)\BM{n}-\nabla_{\bf n}\BM{f}_{0}\right]\,,
\ee
where $\nabla_{\bf n}\equiv(\BM{n},\bn)$.
Equations (\ref{r1.5}) and  (\ref{E4.2}) give
\be\n{f3}
\mbox{div}\BM{f}_{0}=i\sigma[(\BM{f}_{0},\mbox{curl}\,\BM{n})-(\BM{n},\mbox{curl}\BM{f}_{0})]\,.
\ee
and Eqs. (\ref{r1.4}) and (\ref{E4.16}) give
\be\n{f4}
(\BM{f}_{0},\bn)\BM{n}=-[\BM{n}\times\mbox{curl}\BM{f}_{0}]-[\BM{f}_{0}\times\mbox{curl}\,\BM{n}]-\nabla_{\bf n}\BM{f}_{0}\,.
\ee
Substituting the expressions (\ref{f3}) and (\ref{f4}) into (\ref{f2}) and using the relation (\ref{E4.14}) we derive the derivative of $\BM{f}_{0}$ along $\BM{n}$
\be\n{f5}
\nabla_{\bf n}\BM{f}_{0}=\frac{1}{2}\left[i\sigma\BM{n}(\BM{f}_{0},\mbox{curl}\,\BM{n})-[\BM{f}_{0},\mbox{curl}\,\BM{n}] -\BM{f}_{0}\mbox{div}\,\BM{n}\right]\,.
\ee
Taking the scalar product of this expression and the complex conjugate $\bar{\BM{f}}_{0}$ and using Eq. (\ref{r1}) we derive
\be\n{f6}
\nabla_{\bf n}f_{0}^{2}+f_{0}^{2}\mbox{div}\,\BM{n}=\mbox{div}(f_{0}^{2}\BM{n})=0\, .
\ee
This expression implies conservation of number of photons in a stationary spacetime.

Using Eqs. (\ref{f7}), (\ref{f5}), and (\ref{f6}) we obtain the following relation:
\be
i\BM{m} \nabla_{\bf n}\varphi=
\frac{1}{2}\left(i\sigma\BM{n}(\BM{m},\mbox{curl}\,\BM{n})-[\BM{m},\mbox{curl}\,\BM{n}]\right)-\nabla_{\bf n}\BM{m}\, .\n{f10}
\ee
Analogously, we can define derivation of the vector $\BM{n}$ along itself. Using Eqs. (\ref{r1.4}) and (\ref{E4.7}) we have
\be\n{f11}
\nabla_{\bf n}\BM{n}=-[\BM{n}\times\mbox{curl}\,\BM{n}]=[\BM{n}\times\mbox{curl}\,\BM{g}]\,.
\ee
Then, using Eqs. (\ref{E4.10b}), (\ref{gfc}), (\ref{f10}), and (\ref{f11}) we derive
\be\n{f14}
\nabla_{\bf n}\varphi^{\sigma}=\frac{\sigma}{2}(\BM{n},\mbox{curl}\,\BM{g})\,.
\ee
Solving this equation we derive
\be\n{f15}
\varphi^{\sigma}=\sigma \varphi\hh \varphi=\frac{1}{2}\int_{\BM{x}_{0}}^{\BM{x}}(\mbox{curl}\,\BM{g},d\BM{x})\, .
\ee
As above, the integral is evaluated along a ray trajectory connecting a point on $\Gamma$ and a point $\BM{x}$. As a result, solving the transport
equation along the 2-dimensional family of rays determined by the Lagrange manifold one finds the phase $\varphi(\BM{x})$.

To summarize, the application of the `standard' geometrical optics gives the following expression for the vector field ${\cal F}^{\sigma}$:
\ba\n{gen}
{\cal F}^{\sigma}&\approx & f_0^{\sigma}\, \BM{m}^{\sigma} e^{i\omega\tilde{{\cal S}}({\bf x})}\, ,\\
\tilde{{\cal S}}(\BM{x})&=&\int_{\BM{x}_0}^{\BM{x}}\left[1+\left(\tilde{\BM{g}},\frac{d\BM{x}}{d\ell}\right)\right]d\ell\, ,\n{eff}\\
\tilde{\BM{g}}&=& \BM{g}+\frac{\sigma}{2\omega}\mbox{curl}\,\BM{g}\,,\n{ggg}
\ea
where we restored the superscript $\sigma$.

To finish this section consider a superposition of the left and right circularly polarized light beams of the same amplitude
\be
\BM{\cal F}=\BM{\cal F}^{+}+\BM{\cal F}^{-}\, .
\ee
This field describes  linearly polarized beam of light. Using the expression (\ref{gen}) we find
\ba
&&\BM{\cal F}\approx\sqrt{2} f_0\BM{k}_{0}e^{i\omega {\cal S}}\,,\n{f17a}\\
&&\BM{k}_{0}\equiv\BM{e}_{1}\cos\varphi-\BM{e}_{2}\sin\varphi\hhh (\BM{k}_{0},\BM{k}_{0})=1\,.\n{f17b}
\ea
Here $\varphi$ is determined along each of the ray by the relation (\ref{f15}). These relations show that the vector of the linear polarization $\BM{k}_{0}$
rotates with respect to the Fermi propagated basis with the angular velocity
\be\n{frot}
{d\varphi\over d\ell}={1\over 2}( \mbox{curl}\,\BM{g},\BM{n})\, .
\ee
This is the well-known gravitational analogue of the Faraday rotation (see, e.g., \cite{Zo:99,Se:04,CaFeLiRu,FayLlo}).

Let us mention that applying similar steps as above to the higher-order terms in Eq. (\ref{E4.1b}) one can calculate the post-geometrical optics corrections to the obtained solution.

\section{Modified geometrical optics}

\subsection{Modified eikonal equation and  ray trajectories}

We already mentioned that in the `standard' geometrical optics the eikonal equation does not depend on the helicity. As a result, in such a formalism photons of different helicity propagate identically. The spin of a photon experiences a Faraday rotation, but an expected back-reaction of the spin on the photon's trajectory is absent. We already discussed this general problem in Sec. II. The origin of this problem is that the `standard' geometrical optics is valid locally. However, to
desribe the spin effects for photon scattering one needs to modify this approach. For this purpose one needs to include the lowest order helicity corrections into the eikonal equation. Such a modification does not change the local results, and therefore is in agreement with the `standard' geometrical optics locally. However, at large distances,
due to the accumulation of these small corrections,  the trajectories of photons of different helicity are modified considerably.

Equation (\ref{gen}) gives us hints how to made the required modification. In this representation the  equation for the change of the amplitude $f_0$ along the ray trajectory does not depend on helicity. Moreover, the equation for the Fermi transport does not depend on helicity either. The only helicity-dependent term is the phase $\varphi^{\sigma}$. But it is naturally combined with the geometrical optical phase ${\cal S}$ [see the relations (\ref{eff})-(\ref{ggg})].
In a general case, there is an ambiguity in the interpretation of such a phase factor. It is evident that the form of Eq. (\ref{gen}) remains the same if one makes the following transformation:
\be
\BM{f}\to \BM{f} e^{i\psi}\hh {\cal S}\to {\cal S}-\omega^{-1}\psi\, ,
\ee
of the vector amplitude $\BM{f}$ and the phase ${\cal S}$.

The idea of the modified geometrical optics is to fix this ambiguity by imposing the condition that the transport equations for the vector amplitude $\BM{f}$, at least  in the leading order, do not depend on helicity. This has a very non-trivial consequences. Namely, in order to reproduce the expression for $\tilde{{\cal S}}({\bf x})$ in this gauge, one needs to modify the eikonal equation. As a result, the effective Hamiltonian for light rays will be also modified by helicity-dependent corrections. The corresponding Lagrange manifold for our problem will be also modified. At small distances this modification is small, but at large distances the corresponding corrections become important. Since the effective Hamiltonian for light rays and the Largrange manifold depend on helicity, they contain information about the helicity influence on photon trajectories. As expected, the corresponding helicity-dependent terms are suppressed by the small factor $\omega^{-1}$.

After these general remarks let us describe the modified geometrical optics equations in detail.
Let us rewrite the main equation (\ref{E3}) by adding the term ${1\over 2}\omega^{-1}[\mbox{curl}\,\BM{g}\times\BM{f}]$ to its both sides.
The form of this modification is suggested by the effective eikonal (\ref{eff}), (\ref{ggg}),
\be\n{f18}
L\BM{f}+\frac{i}{2\omega}[\mbox{curl}\,\BM{g}\times\BM{f}
]=\frac{\sigma}
{\omega}\mbox{curl}\BM{f}+\frac{i}{2\omega}[\mbox{curl}\,\BM{g}\times\BM{f}]\,.
\ee
Accordingly, we define the following new linear operator $\tilde{L}$ [cf. Eq. (\ref{E4})]:
\ba
&&\tilde{L}\BM{f}\equiv\BM{f}-i\sigma [\tilde{\BM{n}}\times \BM{f}]=0\,,\n{f19a}\\
&&\tilde{\BM{n}}\equiv \BM{p}-\tilde{\BM{g}}\hhh \BM{p}\equiv\bn\bar{S}\hhh \tilde{\BM{g}}\equiv\BM{g}+\frac{\sigma}{2\omega}\mbox{curl}\,\BM{g}\,.\n{f19b}
\ea
Then, Eq. (\ref{f18}) takes the form
\be\n{f20}
\tilde{L}\BM{f}=\frac{\sigma}{\omega}\mbox{curl}\BM{f}+\frac{i}{2\omega}[\mbox{curl}\,\BM{g}\times\BM{f}
]\,.
\ee
Substituting the expansion (\ref{E4.1}) into this equation and taking the geometrical optics limit $\omega^{-1}\to 0$  in accordance with the large distances asymptotic approximation, i.e., we keep the term of the first order in $\omega^{-1}$ which enters the operator $\tilde{L}$, we derive [cf. Eq. (\ref{E4.2})]
\be\n{f21}
\tilde{L}\tilde{\BM{f}}_{0}=0\,.
\ee

One can see that the operator $\tilde{L}$ has the same properties as the operator $L$. In particular, we have [cf. Eq. (\ref{E4.7})]
\be\n{f22}
(\tilde{\BM{n}},\tilde{\BM{n}})=1\,,
\ee
where $\tilde{\BM{n}}$ is the eigenvector of $\tilde{L}$ with the eigenvalue 1.
According to the definition of the vector $\tilde{\BM{n}}$, the effective eikonal equation reads [cf. Eq. (\ref{E7})]
\be\n{f23}
(\bn\tilde{{\cal S}}-\tilde{\BM{g}})^2=1\,.
\ee

This modified eikonal equation can be solved as above. It coincides with the Hamilton-Jacobi equation for the following modified Hamiltonian:
\be\n{modh4}
\tilde{{\cal H}}(x^i,p_i)=\frac{1}{2}(\BM{p}-\tilde{\BM{g}})^2\equiv \frac{1}{2}\gamma^{ij}(p_{i}-\tilde{g}_{i})(p_{j}-\tilde{g}_{j})\, ,
\ee
The modified Hamilton equations are equivalent to the following second order equation for photon trajectories [cf. Eq. (\ref{h8})]:
\be\n{f25}
{D^2\BM{x}\over d\ell^2}=\left[\frac{d\BM{x}}{d\ell}\times\mbox{curl}\,\BM{g}\right]+\frac{\sigma}{2\omega}\left[\frac{d\BM{x}}{d\ell}\times\mbox{curl}\,\mbox{curl}\,\BM{g}\right]\,.
\ee
This equation explicitly shows the effect of the helicity on the  ray trajectories.
Repeating the steps of Sec. IV for $\BM{g}$ replaced with $\tilde{\BM{g}}$ we derive the effective eikonal [cf. Eq. (\ref{eff})]
\be\n{f24}
\tilde{{\cal S}}(\BM{x})=\int_{\BM{x}_0}^{\BM{x}}\left[d\ell+(\BM{g},d\BM{x})+\frac{\sigma}{2\omega}(\mbox{curl}\,\BM{g},d\BM{x})\right]\,.
\ee
The main difference with the previous result is that the integration in (\ref{f24}) is performed over the modified ray trajectories corresponding to the vector $\tilde{\BM{n}}$.
The last term in this equation depends on the helicity sign $\sigma$.

\subsection{Modified transport equations}

Let us now discuss modifications of the transport equations. To begin with, let us introduce the basis $\{\tilde{\BM{n}},\tilde{\BM{m}},\bar{\tilde{\BM{m}}}\}$ such that the relations between the basis vectors are analogous to the relations (\ref{E4.10a}) and (\ref{E4.10b}). Analogously, we define [cf. Eq. (\ref{f7})]
\be\n{f26}
\tilde{\BM{f}}_{0}=\tilde{f}_{0}\tilde{\BM{m}}e^{i\tilde{\varphi}}\,,
\ee
so that
\be\n{f27}
(\tilde{\BM{n}},\tilde{\BM{f}} _{ 0})=0\,.
\ee
From the relation (\ref{E4.13}), which remains valid for the operator $\tilde{L}$ and the vector $\tilde{\BM{n}}$, and Eq. (\ref{f20}) we derive the following relation:
\ba
&&\mbox{curl}\BM{f}=\tilde{\BM{n}}(\tilde{\BM{n}},\mbox{curl}\BM{f})-i\sigma[\tilde{\BM{n}}\times\mbox{curl}\BM{f}]-\frac{i\sigma}{2}[\mbox{curl}\,\BM{g}\times\BM{f}]\nn
&&-\frac{i\sigma}{2}\tilde{\BM{n}}(\mbox{curl}\,\BM{g},[\tilde{\BM{n}}\times\BM{f}])+\frac{1}{2}(\tilde{\BM{n}},\BM{f})\mbox{curl}\,\BM{g}-\frac{1}{2}(\tilde{\BM{n}},\mbox{curl}\,\BM{g})\BM{f}\,.\nn\n{f28}
\ea
Applying the operator $\mbox{curl}$ to Eq. (\ref{f20}) and using Eq. (\ref{f19a}) we derive
\be\n{f29}
\mbox{curl}\BM{f}-i\sigma\mbox{curl}[\tilde{\BM{n}}\times\BM{f}]=\frac{\sigma}{\omega}\mbox{curl}\,\mbox{curl}\BM{f}+\frac{i}{2\omega}\mbox{curl}[\mbox{curl}\,\BM{g}\times\BM{f}]\,.
\ee
Substituting the expansion (\ref{E4.1}) into this equation we derive the following equation corresponding to the geometrical optics approximation [cf. Eq. (\ref{f1a})]:
\be\n{f30}
\mbox{curl}\tilde{\BM{f}}_{0}=i\sigma\mbox{curl}[\tilde{\BM{n}}\times\tilde{\BM{f}}_{0}]\,.
\ee
Repeating the same steps as in Sec. IV and using Eqs. (\ref{f26}), (\ref{f27}), and the relation (\ref{f28}) we derive [cf. Eq. (\ref{f5})]
\ba
&&\nabla_{\tilde{{\bf n}}}\tilde{\BM{f}}_{0}=\frac{1}{2}\left[i\sigma\tilde{\BM{n}}(\tilde{\BM{f}}_{0},\mbox{curl}\,\tilde{\BM{n}})-[\tilde{\BM{f}}_{0},\mbox{curl}\,\tilde{\BM{n}}] -\tilde{\BM{f}}_{0}\mbox{div}\,\tilde{\BM{n}}\right]\nn
&&-\frac{i\sigma}{4}\left[\tilde{\BM{n}}(\tilde{\BM{f}}_{0},\mbox{curl}\,\BM{g})-i\sigma[\tilde{\BM{f}}_{0},\mbox{curl}\,\BM{g}] +\tilde{\BM{f}}_{0}(\tilde{\BM{n}},\mbox{curl}\,\BM{g})\right]\,.\nn\n{f31}
\ea
Using this expression, one can show that the following relation holds [cf. Eq. (\ref{f6})]:
\be\n{f32}
\mbox{div}(\tilde{f}_{0}^{2}\tilde{\BM{n}})=0\,.
\ee
That is, conservation of number of photons in a stationary spacetime holds for the modified ray trajectories.
Using the expressions (\ref{f26}), (\ref{f31}), and (\ref{f32}) we obtain the following relation [cf. Eq. (\ref{f10})]:
\ba
i\tilde{\BM{m}} \nabla_{\tilde{{\bf n}}}\tilde{\varphi}&=&\frac{1}{2}\left(i\sigma\tilde{\BM{n}}(\tilde{\BM{m}},\mbox{curl}\,\tilde{\BM{n}})-[\tilde{\BM{m}},\mbox{curl}\,\tilde{\BM{n}}]\right)\nn
&-&\frac{i\sigma}{4}\left[\tilde{\BM{n}}(\tilde{\BM{m}},\mbox{curl}\,\BM{g})-i\sigma[\tilde{\BM{m}},\mbox{curl}\,\BM{g}] \right.\nn
&+&\left.\tilde{\BM{m}}(\tilde{\BM{n}},\mbox{curl}\,\BM{g})\right]-\nabla_{\tilde{{\bf n}}}\tilde{\BM{m}}\,.\n{f33}
\ea
Analogously, using Eqs. (\ref{r1.4}), (\ref{f19b}), and (\ref{f22}) we can define derivation of the vector $\tilde{\BM{n}}$ along itself [cf. Eq. (\ref{f11})],
\be\n{f34}
\nabla_{\tilde{{\bf n}}}\tilde{\BM{n}}=-[\tilde{\BM{n}}\times\mbox{curl}\,\tilde{\BM{n}}]=[\tilde{\BM{n}}\times\mbox{curl}\,\BM{g}]+\frac{\sigma}{2\omega}[\tilde{\BM{n}}\times\mbox{curl}\,\BM{g}]\,.
\ee
We use the basis $\{\tilde{\BM{n}},\tilde{\BM{m}},\bar{\tilde{\BM{m}}}\}$, which is Fermi propagated along the corresponding ray trajectories generated by $\tilde{\BM{n}}$. Then, using Eqs. (\ref{f33}) and (\ref{f34}) we derive [cf. Eq. (\ref{f14})]
\be\n{f35}
\nabla_{\tilde{{\bf n}}}\tilde{\varphi}=\frac{1}{4\omega}(\tilde{\BM{n}},\mbox{curl}\,\mbox{curl}\,\BM{g})\,.
\ee
Thus, in the geometrical optics limit $\omega^{-1}\to0$ we have $\tilde{\varphi}=const$, which implies that polarization vector does not rotate with respect to the basis $\{\tilde{\BM{n}},\tilde{\BM{m}},\bar{\tilde{\BM{m}}}\}$.

To summarize, the expression for the field $\BM{{\cal F}}^{\sigma}$ in the modified geometrical optics approximation reads
\ba\n{modgen}
{\cal F}^{\sigma}&\approx & \tilde{f}_0^{\sigma}\, \tilde{\BM{m}}_{\sigma} e^{i\omega\tilde{{\cal S}}({\bf x})}\, ,\\
\tilde{{\cal S}}({\bf x})&=&\int_{\BM{x}_0}^{\BM{x}}\left[1+\left(\tilde{\BM{g}},\frac{d\BM{x}}{d\ell}\right)\right]d\ell\, ,\n{modeff}\\
\tilde{\BM{g}}&=& \BM{g}+\frac{\sigma}{2\omega}\mbox{curl}\,\BM{g}\, .\n{modggg}
\ea
This result is similar to (\ref{gen}), but it contans an important difference. To define the Fermi propagated basis and the integration paths one must use the modified ray trajectories defined by the tangent vector $\tilde{\BM{n}}$.

Since a linearly polarized light is a superposition of the right and left circularly polarized beams, we conclude that the initially linearly polarized beam in a stationary gravitational field splits into two beams with different circular polarization.
However, it is easy to check that before this effect becomes important the beam split is negligible, and the linear polarizion vector has the same Faraday rotation (\ref{frot}).

\subsection{Example: Kerr spacetime}

Let us emphasize that in the developed scheme of the modified geometrical optics we did not assume that the gravitational field is weak. One can expect that spin-optical effects might become significant for the light propagation close to rotating black holes, where the gravity is evidently very strong. We shall discuss this subject in detail in the future publication. Just to give an example, here we present the terms which control the helicity dependence of light rays in the Kerr geometry.

The Kerr metric in the Boyer-Linquist coordinates $x^{\mu}=(t,r,\theta,\phi)$  can be presented in the form (\ref{A.57}) with
\ba
&&\hspace{-0.9cm}h=(\Delta-a^{2}\sin^{2}\theta)/\Sigma\hhh g_{i}=-\frac{2aMr}{\Sigma h}\sin^{2}\theta\delta_{i}^{~\phi}\,,\n{K1}\\
&&\hspace{-0.9cm}d\ell^{2}=\gamma_{ij}dx^{i}dx^{j}=\frac{\Sigma}{\Delta h}dr^{2}+\frac{\Sigma}{h}d\theta^{2}+\frac{\Delta\sin^{2}\theta}{h^{2}} d\phi^{2}\,,\n{K2}
\ea
where
\be\n{K3}
\Sigma=r^{2}+a^{2}\cos^{2}\theta\hhh \Delta=r^{2}-2Mr+a^{2}\,.
\ee
This metric represents a Kerr black hole of the mass $M$ and the
angular momentum $J=aM$, where $0\leq|a|\leq M$.

For this metric we derive
\ba
&&\hspace{-1.0cm}(\mbox{curl}\,\BM{g})^{i}=-\frac{4aM\Delta r}{\Sigma^{3}}\cos\theta\delta^i_{~r}\nn
&&\hspace{0.3cm}-\frac{2aM(r^{2}-a^{2}\cos^{2}\theta)}{\Sigma^{3}}\sin\theta\delta^i_{~\theta}\,,\n{K4a}\\
&&\hspace{-1.0cm}(\mbox{curl}\,\mbox{curl}\,\BM{g})^{i}=\frac{4aM^{2}}{\Sigma^{3}}\delta^i_{~\phi}\,.\n{K4b}
\ea
Note that Eq.  (\ref{f25}) for photon trajectories contains an extra term (\ref{K4b}) which in the weak field approximation gives the angular separation of the right and left circularly polarized beams of the order of $GJ\lambda/(c^{3}L^{3})$, where $L\gg GM/c^{2}$ is the distance from the Kerr black hole. This result is in agreement with the estimation obtained by Mashhoon \cite{Ma:75,Ma:74a}.

\section{Summary}

In this paper we studied the propagation of circularly polarized monochromatic electromagnetic waves in a stationary spacetime. We presented the Maxwell equations in (3+1)-form corresponding to (3+1)-decomposition of a stationary spacetime. The Maxwell equations for circularly polarized monochromatic waves reduce to the set of equations for the complex amplitude of such waves. This set of equations can be solved if a solution to the corresponding master equation for the complex amplitude is found. We first described the `standard' procedure of finding solutions to the master equation in the geometrical optics approximation. We used the main result of this approximation (\ref{gen}) in order to develop an improved modified scheme. This scheme locally gives the same results as the `standard' geometrical optics, but
for the scattering problem it allows one to include the helicity corrections to the photon equation of motion. It is interesting that technically, in order to obtain these modified equations, it is sufficient to modify the gravitomagnetic vector potential $\BM{g}$ by the well-defined correction which has the form ${1\over 2}\sigma \omega^{-1}\mbox{curl}\,\BM{g}$. Let us emphasize that the obtained results are valid beyond the weak field approximation for the gravitational field, and hence, they allow one to study the polarization effects in strong gravity. Using the proposed scheme one can calculate the post-geometrical optics corrections to the obtained solution.

One can interprete the obtained results in a slightly different manner. As a result of spin-orbit interaction, photons with different circilar polarization move in a slightly modified metric. This modification depends on the frequency of a photon $\omega$. Such `renormalization' approach allows one to conclude that due to the scattering of circularly polarized `white' light, which contains different frequencies, one may observe a peculiar `rainbow effect'. Another effect is that  linearly polarized beam of light propagating in a stationary gravitational field splits into two components of the right and left circular polarization.
These and other polarization effects might be important for propagation of polarized photons near the horizon of rapidly rotating black holes.

\begin{acknowledgments}

This work was partly
supported  by  the Natural Sciences and Engineering Research Council
of Canada. One of the authors (V. F.) is grateful to the Killam Trust
for its financial support.
\end{acknowledgments}

\appendix

\section{Calculation of determinant}

Suppose that $\alpha$ is some scalar quantity, $\BM{I}$ is a unit and
$\BM{A}$ is an antisymmetric $3\times 3$ matrices, respectively. Then one has
\be
\mbox{det}(\alpha\BM{I} +\BM{A})= \alpha^3 -\frac{\alpha}{2}\mbox{tr}(\BM{A}^{2})\,.
\ee
The antisymmetric matrix $\BM{A}$ can be written in the form
\be
A_{ij}= \epsilon_{ijk} a^k\, ,
\ee
where $\epsilon_{123}=\epsilon^{123}=1$ and $a^i={1\over 2} \epsilon^{ijk} A_{jk}$.
Using this representation one
obtains
\be
\mbox{tr} (\BM{A}^2)=-2 \BM{a}^2=-2 a_i a^i\,.
\ee
Thus,
\be
\mbox{det}(\alpha\BM{I} +\BM{A})= \alpha^3 +\alpha a_i a^i\,.
\ee

\end{document}